\documentclass[prl,twocolumn,showpacs,superscriptaddress]{revtex4}
\usepackage{graphicx,psfrag,amsmath,amssymb,amsfonts,bbm,latexsym}

\newcommand{\ve}{\ensuremath{\varepsilon}}

\begin{document}
\title{Quantum breaking time near classical equilibrium points} 
\author{Fabrizio Cametti}
\affiliation{Dipartimento di Fisica, Universit\`a di Roma ``La Sapienza'',
Piazzale A. Moro 2, Roma 00185, Italy}
\author{Carlo Presilla}
\affiliation{Dipartimento di Fisica, Universit\`a di Roma ``La Sapienza'',
Piazzale A. Moro 2, Roma 00185, Italy}
\affiliation{Istituto Nazionale di Fisica Nucleare, Sezione di Roma 1} 
\affiliation{Istituto Nazionale per la Fisica della Materia, 
Unit\`a di Roma 1 and Center for Statistical Mechanics and Complexity} 

\date{January 11, 2002}

\begin{abstract}
By using numerical and semiclassical methods, we evaluate the quantum
breaking, or Ehrenfest time for a wave packet localized around
classical equilibrium points of autonomous one-dimensional systems
with polynomial potentials. We
find that the Ehrenfest time diverges logarithmically with the inverse
of the Planck
constant whenever the equilibrium point is exponentially unstable.
For stable equilibrium points, we have a power law divergence with
exponent determined by the degree of the potential near the
equilibrium point.  
\end{abstract}
\pacs{03.65.Sq,05.45.Mt,47.52.+j}
\maketitle

The question of estimating how long classical and quantum evolutions stay
close is one of the main problems of semiclassical analysis. The
evolution of a quantum observable can follow that of the corresponding  
classical one up to a finite time, the so called quantum breaking, or
Ehrenfest, time. As initially conjectured in \cite{BZ,Z} and
rigorously proved in \cite{CR,BGP,HJ}, whenever the classical flow is
chaotic, the Ehrenfest time diverges logarithmically in $\hbar$. 
This result is easily understood. Starting from an initial value
$\Delta(\hbar)\sim \hbar/I$, where $I$ is a characteristic action of
the system, the difference between a
classical flow, with Lyapunov exponent $\lambda>0$, and
the corresponding quasi-periodic quantum flow increases as
$\Delta(\hbar)\exp(\lambda t)$. The two flows depart at $t\sim
\lambda^{-1}\log (I/\hbar)$. The situation is different for a regular
classical flow. In this case, 
starting from the work \cite{FGP}, it was suggested in \cite{LOG} that
the Ehrenfest 
time grows algebraically as $\hbar^{-\delta}$. The determination of
the value of $\delta$ and its possible universal nature is still an
open problem. See \cite{IZ} and references therein for recent results.  

The $\hbar$-scaling of the Ehrenfest time is usually
investigated 
for classical flows which are completely chaotic or
regular. 
However,
it is interesting to study the quantum-classical correspondence in
systems having isolated unstable orbits embedded in a completely
regular phase-space.
The simplest example is given by the ubiquitous double-well system
defined by the Hamiltonian $H(p,q) =\frac{p^2}{2}-\frac{q^{2}}{2}
+\frac{q^{4}}{4}$.   
For this system, there is only one unstable periodic orbit, namely
that associated to the equilibrium point $(p_0,q_0)=(0,0)$, with positive
Lyapunov exponent $\lambda = 1$. Is it possible to have a logarithmic
Ehrenfest time in proximity of an isolated exponentially unstable point 
like $(p_0,q_0)$?  

The usual way of studying the Ehrenfest time consists in comparing the
evolution of classical observables with the quantum expectation value
of the corresponding operators, either in the coherent state
representation \cite{CR}, or in the framework of Weyl
quantization \cite{BGP}.  
In the present case, we follow a simpler approach based on the
analysis of the 
quantum spectrum. We know that on going towards the classical 
equilibrium point $(p_0,q_0)$ the period of motion diverges, so that
the evolution of a phase-space distribution function localized around
this point must show a continuous frequency distribution around
$\nu=0$.    
On the other hand, in the quantum case, due to the
discrete nature of the spectrum, the frequency distribution is
characterized by a gap between zero and a minimal frequency. We call
this minimal frequency the Ehrenfest frequency, $\nu_E$. In fact, its
inverse, $\nu_E^{-1}$, is an upper bound to the time at which the
quantum-classical correspondence of the evolution of any observable
breaks down. We estimate the Ehrenfest time as $\nu_E^{-1}$.

By using numerical and semiclassical methods, we study the behavior of
$\nu_E(\hbar)$ around classical equilibrium points, both stable and
unstable, for several autonomous one-dimensional systems. 
We find that $\nu_E^{-1}(\hbar)$ diverges logarithmically for $\hbar
\to 0$ whenever the
equilibrium point is exponentially unstable.
In all the other cases, the Ehrenfest time follows a power law with
exponent related to the degree of the potential near the
equilibrium point. 
 
In the following, we consider systems described by the Hamiltonians
\begin{equation}
H(p,q) =\frac{p^2}{2m}+ A \frac{q^{2\alpha}}{2\alpha} + B 
\frac{q^{2\beta}}{2\beta},
\label{hnr}
\end{equation}
with $A \le 0$, $B>0$ and $\beta>\alpha\ge 1$.
By properly rescaling position, momentum and time, we can always
reduce to the case $B=1$, $m=1$ and either $A=0$ or $A=-1$ \footnote{%
For $A=0$, the physical energy and the Planck constant are given in terms
of the 
corresponding rescaled quantities via the substitution 
$\ve \to
\ve~ m^{-\frac{\beta}{\beta-1}} B^{\frac{1}{\beta-1}}
\tau^{\frac{2\beta}{\beta-1}}$,    
$\hbar \to \hbar~ m^{-\frac{\beta}{\beta-1}} B^{\frac{1}{\beta-1}}
\tau^{\frac{\beta+1}{\beta-1}}$,
where $\tau$ is an arbitrary time scale unit.
In the double-well case, $A<0$, we have
$\ve \to \ve~ (-A)^{-\frac{\beta}{\beta-\alpha}}
B^{\frac{\alpha}{\beta-\alpha}}$  and
$\hbar \to
\hbar~ m^{-\frac{1}{2}} (-A)^{-\frac{\beta+1}{2(\beta-\alpha)}}
B^{\frac{\alpha+1}{2(\beta-\alpha)}}$.}.
For $A=0$, we have single-well systems with a classical stable
equilibrium point 
$(p_0,q_0)=(0,0)$ at energy $\ve=0$. 
A more interesting situation occurs for $A=-1$. In this case the
systems are
double-well oscillators and the classical equilibrium point
$(p_0,q_0)=(0,0)$ at 
energy $\ve=0$ is unstable. 
In the particular 
case $\alpha=1$, the equilibrium point is exponentially unstable.
In both cases, $A=0$ or $A=-1$, the periodic orbits near the
equilibrium point at $\ve=0$ have a period which diverges for $\ve \to
0$.
\begin{figure}[t]
  \begin{center}
    \psfrag{nu}{$\nu$}
    \psfrag{P}{$\mathcal{P}(\nu)$}
    \psfrag{h1}{\hspace{-2mm}$\hbar=10^{-4}$}
    \psfrag{h2}{\hspace{-2mm}$\hbar=10^{-8}$}
    \psfrag{h3}{\hspace{-2mm}$\hbar=10^{-12}$}
    \includegraphics[width=0.90\columnwidth]{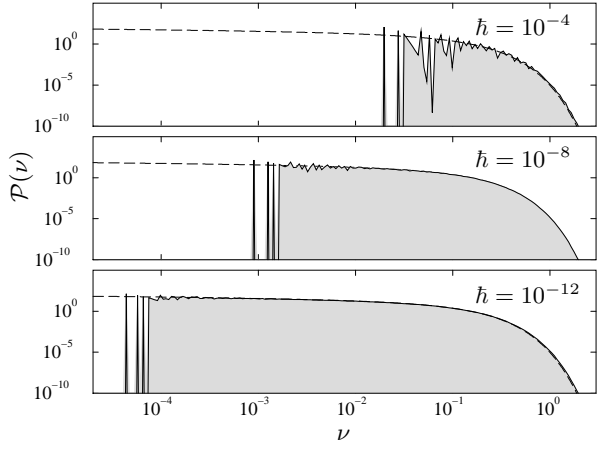}
    \caption{Fourier transform of the survival probability
    $\mathcal{P}(\nu)$ for different values of $\hbar$ in the case
    $A=0$, $\beta=2$. The dashed line is the $\hbar \to 0$ limit
    distribution given by Eq.~(\ref{Plimit}).} 
    \label{fig1}
  \end{center}
\end{figure}

On the quantum mechanical side, in order to represent a state
localized near the classical equilibrium point $(p_0,q_0)$ we consider the
following initial wavefunction
\begin{equation}
\label{iwf}
\langle q|\psi(0)\rangle = \frac{1}{(\pi \hbar)^{\frac{1}{4}}}\exp \left[
    -\frac{(q-q_0)^2}{2\hbar}\right] \exp \left(i\frac{p_0
q}{\hbar}\right).    
\end{equation}
The associated Wigner function,
\begin{eqnarray}
W_\psi(p,q)=\frac{1}{\pi\hbar} \exp\left[-\frac{(p-p_0)^2}{\hbar}\right]
\exp\left[-\frac{(q-q_0)^2}{\hbar}\right], 
\end{eqnarray}
can be interpreted as a phase-space
probability distribution centered around the point $(p_0,q_0)$ 
and has the property
\begin{eqnarray}
\lim_{\hbar \to 0} W_\psi(p,q) = \delta(p-p_0)\delta(q-q_0).
\end{eqnarray}
In these expressions $\hbar$ is the adimensional rescaled
Planck constant, which vanishes when, for instance, the mass $m$
of the system is taken larger and larger.   

Instead of considering the evolution of a specific observable, we
study the simpler survival probability 
\begin{equation}
 \mathcal{P}(t) = \vert\langle\psi(0)|\psi(t)\rangle\vert^2,
\end{equation}
which contains the same gross dynamical information.
In the basis of the eigenstates of the Hamiltonian,
\begin{equation}
H |\phi_n\rangle = \ve_n |\phi_n\rangle,\quad n=0,1,2,\ldots,
\end{equation}
the survival probability $\mathcal{P}(t)$ can be written as
\begin{eqnarray}
  \label{sp}
  \mathcal{P}(t) &=& \sum_{n=0}^\infty\sum_{m=0}^\infty |c_n|^2
  |c_m|^2 \exp 
    \left(i\nu_{nm} t\right),
\end{eqnarray}
where $c_k = \langle \psi(0) | \phi_k\rangle$ and
$\nu_{nm}=(\ve_n-\ve_m)/(2\pi \hbar)$. Note that $c_k=0$ for $k$ odd,
due to the symmetry of the system and of the initial wavefunction. 

By using semiclassical and numerical techniques, we now show that 
the Fourier transform of the
survival probability, 
\begin{eqnarray}
  \label{tfsp}
  \mathcal{P}(\nu) &=& \sum_{n=0}^\infty\sum_{m=0}^\infty |c_n|^2
  |c_m|^2 \delta(\nu-\nu_{nm}),
\end{eqnarray}
for sufficiently small values of $\hbar$
is characterized by a gap, large with respect to the typical level
spacing, between $\nu=0$ and a frequency which we call the Ehrenfest
frequency, defined as
\begin{eqnarray}
\label{nue}
\nu_E=\min_{n\ne m \atop |c_n|^2|c_m|^2 \neq 0}\nu_{nm}.
\end{eqnarray}
\begin{figure}[t]
  \begin{center}
    \psfrag{x}{$q$}
    \psfrag{y40}{$\phi_{40}(q)$}
    \psfrag{y42}{$\phi_{42}(q)$}
    \psfrag{e40}{\hspace{-4mm}$\ve_{40}=-0.0054148$}
    \psfrag{e42}{\hspace{-4mm}$\ve_{42}=0.0008148$}
    \includegraphics[width=0.90\columnwidth]{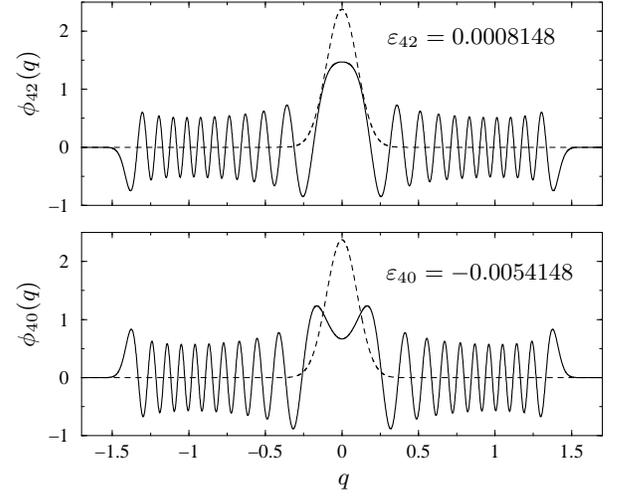}
    \caption{Eigenfunctions $\phi_{40}$ and $\phi_{42}$ corresponding
    to the minimal frequency 
$\nu_E$ in the double-well case $\alpha=1$, $\beta=2$ for 
$\hbar=10^{-2}$. The dashed curve is the initial wavefunction
(\protect\ref{iwf}) with $(p_0,q_0)=(0,0)$.}
    \label{fig2}
  \end{center}
\end{figure}
\begin{figure}[t]
  \begin{center}       
    \psfrag{e}{$\ve_n/\hbar$}
    \psfrag{c}{$|c_n|^2$}       
    \includegraphics[width=0.90\columnwidth]{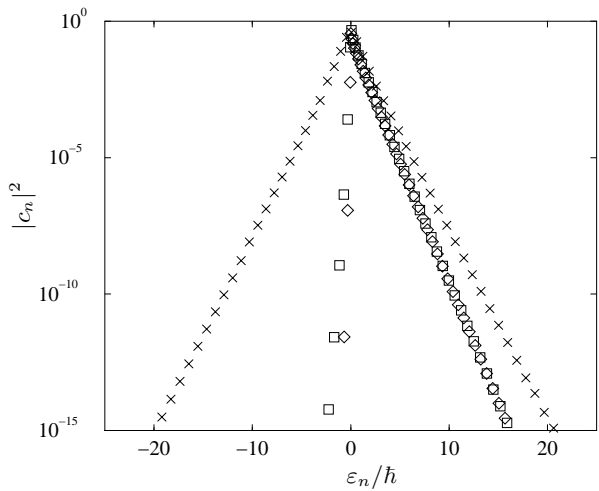}
    \caption{Superposition coefficients $|c_n|^2$ as a function of 
$\ve_n/\hbar$ for $\hbar=10^{-3}$ in the double-well cases 
$\alpha=1$, $\beta=2$ ($\times$),
$\alpha=2$, $\beta=4$ ($\Box$),
and $\alpha=3$, $\beta=6$ ($\Diamond$).}
    \label{fig3}
  \end{center}
\end{figure}

In the simple case $A=0$, by using standard WKB
approximations, we have  
\begin{eqnarray} 
\label{wkbe}
\ve_n =\left[\left(n+\frac{1}{2}\right)\hbar
  \delta(\beta)\right]^{\frac{2\beta}{\beta+1}}, 
\end{eqnarray}
with 
\begin{eqnarray}
\delta(\beta) = \sqrt{\frac{\pi}{2}}\frac{
\Gamma\left[\frac{1}{2}\left(3+\frac{1}{\beta}\right)
\right]}{\Gamma\left(1+\frac{1}{2\beta}\right)  
(2\beta)^{\frac{1}{2\beta}}},
\end{eqnarray}
and
\begin{eqnarray}
\label{wkbc}
|c_\ve|^2 =
\frac{2\sqrt{\pi}
\left(2\beta\right)^{-\frac{1}{2\beta}}\hbar^{\frac{1}{2}}
\ve^{-\frac{1}{2\beta}} e^{-2\frac{\ve}{\hbar}}}
{\frac{\Gamma(\frac{1+\beta}{2\beta})} 
{\Gamma(\frac{1}{2})\Gamma(1+\frac{1}{2\beta})}
+ \frac{\sin \sigma(\ve,\hbar;\beta)}{\sigma(\ve,\hbar;\beta)}},
\end{eqnarray}
with $\sigma(\ve,\hbar;\beta) =
2\sqrt{2}(2\beta)^{\frac{1}{2\beta}}
\hbar^{-1}\ve^{\frac{\beta+1}{2\beta}}$. 
The behavior of $\mathcal{P}(\nu)$ obtained
by using these expressions for $\ve_n$ and $|c_{\ve_n}|^2$ is shown in
Fig.~\ref{fig1} in the case $\beta=2$. 
We see that for $\hbar \to 0$ the frequency distribution
$\mathcal{P}(\nu)$ approaches a continuous limit given by 
\begin{eqnarray}
\mathcal{P}_0(\nu)=\lim_{\hbar \to 0} \int\! \mathrm{d}\ve_1
\mathrm{d}\ve_2\
p(\ve_1)p(\ve_2)\delta\!\left(\nu-\frac{\ve_1-\ve_2}{2\pi \hbar}\right), 
\end{eqnarray}
where $p(\ve)=|c_{\ve}|^2\frac{\mathrm{d} n}{\mathrm{d}\ve}$ and
$n(\ve)$ is obtained by inverting $\ve=\ve_n$. By using (\ref{wkbe}) and
(\ref{wkbc}), we find 
\begin{eqnarray}
\mathcal{P}_0(\nu)=4 K_0(4\pi |\nu|),
\label{Plimit}
\end{eqnarray}
where $K_0$ is the Bessel function of zero-th order.
Figure~\ref{fig1} also shows the presence of the gap at $\nu=0$ and its
shrinking as 
$\hbar \to 0$. 
Since the level spacing $\ve_{n+1}-\ve_{n}$ increases by increasing $n$,
the Ehrenfest frequency (\ref{nue}) turns out to be
$\nu_E=(\ve_2-\ve_0)/(2\pi \hbar)$. According to 
(\ref{wkbe}), its inverse diverges as 
\begin{eqnarray}
\label{ves}
\nu_E^{-1} \sim \hbar^{\frac{1-\beta}{1+\beta}}.
\end{eqnarray}

We now consider double-well systems, i.~e., the case $A=-1$. For these
systems, the standard WKB approximation fails near the unstable
equilibrium point at energy $\ve=0$. Only in the particular case
$\alpha=1$, a regularized semiclassical approximation has been
developed \cite{FFL,cdvp1,cdvp2} and the quantization condition for
the energy levels reads 
\begin{eqnarray}
\label{rwkbe1}
\frac{1}{\sqrt{1+\exp{\frac{2\pi\ve}{\hbar}}}}=\cos(\phi(\ve,\hbar)),
\end{eqnarray}
where 
\begin{eqnarray}
\label{rwkbe2}
\phi(\ve,\hbar)=\frac{4}{3\hbar}-\frac{\ve}{\hbar}\log\frac{\hbar}{16}
- \arg \Gamma\left(\frac{1}{2}+i\frac{\ve}{\hbar}\right) - \pi. 
\end{eqnarray} 
For the associated eigenfunctions only microlocal expressions are 
available \cite{cdvp1} which do not allow for a direct
determination of the superposition coefficients $|c_\ve|^2$. 
For this reason, in all cases 
$\alpha\ge 1$ we determine numerically the eigenvalues and
eigenfunctions of the system. 
With standard numerical techniques, 
this represents an unsurmountable task since the
interesting eigenstates, namely those close to 
energy $\ve=0$, have a quantum number $n$ which diverges quickly for
$\hbar \to 0$. We bypass the problem by using the algorithm \cite{PT}
which allows to evaluate selected eigenstates having a very large
number of nodes.
In Fig.~\ref{fig2} we show, as an example, the couple of even
eigenfunctions 
with energy closest to $\ve=0$ in the double-well case $\alpha=1$,
$\beta=2$ evaluated for $\hbar=10^{-2}$.  
Note that, already for this still relatively large value of $\hbar$, the
corresponding quantum number is $n \sim 40$. In our numerical
calculations we go beyond $n \sim 10^4$.   

In Fig.~\ref{fig3} we show the superposition coefficients evaluated
for different double-well systems for $\hbar= 10^{-3}$. 
We see that $|c_n|^2$ decreases exponentially departing
from $\ve=0$.
For smaller values of $\hbar$, the superposition coefficients
$|c_n|^2$ follow approximately the same exponential behavior as a
function of $|\ve_n|/\hbar$ and become denser and denser.
\begin{figure}[t!]
  \begin{center}
    \psfrag{nu}{$\nu$}
    \psfrag{P}{$\mathcal{P}(\nu)$}
    \psfrag{h1}{\hspace{-2mm}$\hbar=10^{-2}$}
    \psfrag{h2}{\hspace{-2mm}$\hbar=10^{-3}$}
    \psfrag{h3}{\hspace{-2mm}$\hbar=10^{-4}$}
    \includegraphics[width=0.90\columnwidth]{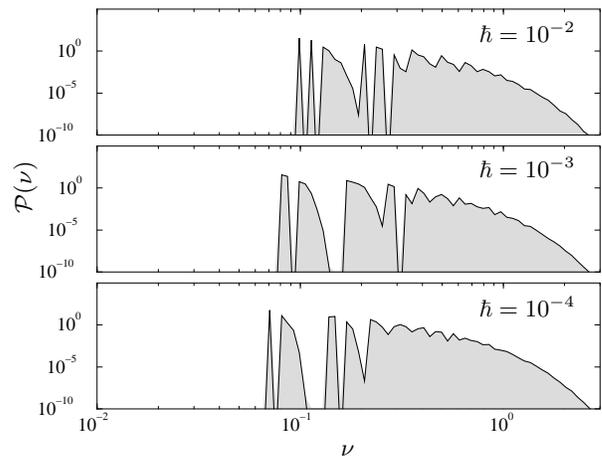}
    \caption{Fourier transform of the survival probability
    $\mathcal{P}(\nu)$ for different values of $\hbar$ in the
    double-well case 
     $\alpha=1$, $\beta=2$.}
    \label{fig4}
  \end{center}
\end{figure}

The Fourier transform of the survival
probability (\ref{tfsp}) is determined by using the eigenvalues and
the superposition coefficients obtained numerically. 
In Fig.~\ref{fig4} we show
$\mathcal{P}(\nu)$ in the case $\alpha=1$, $\beta=2$ for different
values of $\hbar$. As in the single-well case, at $\nu=0$ we have a
gap whose width shrinks as $\hbar \to 0$. The width of this gap,
namely the Ehrenfest frequency, is
yielded by a couple of even consecutive eigenvalues, close to the
energy $\ve=0$ of the 
classical equilibrium point. This can be understood roughly in the
following way. Consider the number of states, $\mathcal{N}_\ve$, in the
energy range $[\ve-\hbar,\ve+\hbar]$. The frequencies associated to
the eigenvalues in this energy range are $\nu \sim
\mathcal{N}_\ve^{-1}$, so that, in the limit $\hbar \to 0$, 
$\nu$ vanishes if $\mathcal{N}_\ve$ diverges.
According to Weyl formula, $\mathcal{N}_\ve$ 
is proportional to the classical phase-space volume bounded by the energy
shells $H(p,q)=\ve\pm\hbar$. 
This volume can be evaluated exactly in terms of simple functions in the
single-well case and in terms of special functions for double-well
systems. In all cases, we have that $\mathcal{N}_\ve$ diverges when
$\hbar \to 0$ only for $\ve=0$. In the double-well systems, 
for $\alpha=1$, $\beta=2$ the couple of closest eigenvalues has 
energies of opposite sign, as shown
in Fig.~\ref{fig2}. For $\alpha > 1$ these 
eigenvalues are both positive if $\hbar$ is sufficiently small.  
\begin{figure}[]
  \begin{center}
    \psfrag{hbar}{$\hbar$}
    \psfrag{1/nu}{$\nu_E^{-1}$}
    \psfrag{a}{\hspace{-4mm}$\nu_E^{-1}\sim \hbar^{-1/3}$}
    \psfrag{c}{\hspace{-5mm}$\nu_E^{-1}\sim \hbar^{-1/2}$}
    \psfrag{b}{$\nu_E^{-1}\sim \log \hbar^{-1}$}
    \includegraphics[width=0.90\columnwidth]{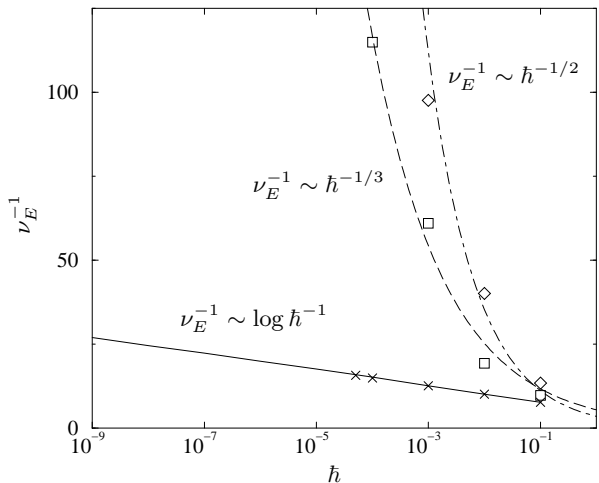}
    \caption{Inverse of the Ehrenfest frequency, $\nu_E^{-1}$, as a
    function of $\hbar$ in the double-well cases 
$\alpha=1$, $\beta=2$ ($\times$), $\alpha=2$, $\beta=4$ ($\Box$),
and $\alpha=3$, $\beta=6$ ($\Diamond$). The solid line is the
    regularized WKB prediction based on (\ref{rwkbe1}-\ref{rwkbe2}),
    while the dashed and dot-dashed lines are numerical fits.} 
    \label{fig5}
  \end{center}
\end{figure}

The scaling of $\nu_E^{-1}$ with $\hbar$ is shown in Fig.~\ref{fig5}
for different double-well systems. The plotted points are calculated
using the numerically determined spectrum while the solid line
represents the inverse of the Ehrenfest frequency as determined by using
the quantization condition
(\ref{rwkbe1}-\ref{rwkbe2}). The Ehrenfest time increases
logarithmically with $\hbar^{-1}$ only in 
the case $\alpha=1$, $\beta=2$, i.~e., when the equilibrium point is
exponentially unstable. In all the other cases, a
numerical fit suggests that
\begin{eqnarray}
\label{ss}
\nu_E^{-1} \sim \hbar^{\frac{1-\alpha}{1+\alpha}}.
\end{eqnarray}
This is the same scaling law which we would obtain, as described by
Eq.~(\ref{ves}), in the case of a
single-well potential $V(q)=q^{2\alpha}/(2\alpha)$. 
This fact can be understood in the following way. 
For $\hbar \to 0$, the discrete eigenvalues of the double-well above
$\ve=0$ 
correspond to the energies of the continuous spectrum of the barrier
$-q^{2\alpha}/(2\alpha)$ at which the transmission
coefficient is maximum. According to WKB approximation, these
resonances of the continuous spectrum in turn coincide 
with the energies of the bound states of the corresponding confining
inverted potential.  

In conclusion, we have shown that the presence of isolated
exponentially unstable orbits is sufficient to break the
quantum-classical correspondence at a time scale logarithmic in
$\hbar^{-1}$. 
This feature may be relevant in all mesoscopic systems which
are modeled by one-dimensional multi-well Hamiltonians
\cite{JLPC,firenze}. In these systems the Ehrenfest time behavior 
is related to experimentally detectable properties as the classical to
quantum crossover of the shot noise \cite{AAL}. 

\begin{acknowledgments}
We would like to thank Thierry Paul for very stimulating discussions.
This research was partially supported by 
Cofinanziamento MURST protocollo MM02263577\_001.  
\end{acknowledgments}
\vspace{0.5cm}

\end{document}